\documentclass{article} % For LaTeX2e
\usepackage[preprint]{tmlr}

\usepackage{amsmath,amssymb,amsthm}
\usepackage[utf8]{inputenc}
\usepackage{graphicx}
\usepackage{booktabs}
\usepackage{hyperref}
\usepackage{xcolor}
\usepackage{microtype}
\usepackage{natbib}
\usepackage{caption}
\usepackage{multirow}
\usepackage{url}
\usepackage{float}

\hypersetup{colorlinks=true, linkcolor=blue!60!black,
            citecolor=blue!60!black, urlcolor=blue!60!black}

\title{How a Cooperative-Override Circuit Suppresses
Nash Play in Large Language Models}

\author{Paraskevas Lekeas\thanks{Corresponding author.} \\
Morton Grove, IL, USA \\
\texttt{plekeas@gmail.com}
\AND
Giorgos Stamatopoulos \\
Department of Economics, University of Crete \\
Rethymno, Crete, Greece \\
\texttt{gstamato@uoc.gr}}

\def\month{MM}
\def\year{YYYY}
\def\openreview{\url{https://openreview.net/forum?id=XXXX}}

\begin{document}
\maketitle

\begin{abstract}
On the named Prisoner's Dilemma under direct prompting, three larger
instruction-tuned models, Llama-3-70B, Qwen2.5-32B, and Qwen2.5-72B,
lock at full cooperation, the metric's maximum distance from Nash with
zero variance across replicates, while Llama-3-8B plays near-Nash. Opening the models, a logit-lens analysis finds a distributed
cooperative override. Intermediate readouts lean toward the Nash action
through roughly three quarters of network depth before a late surge
toward cooperation, and the final layer settles the contest. The size of that final correction, not the
surge, rank-matches chain-of-thought behavior across scale and two
architectures. In the 8B the override is a single causally controllable direction in
the residual stream; steering it dials the decision, and clamping its
component at one position of one layer moves the choice strictly
monotonically, Spearman $\rho = 1.000$, with generation fluent. The circuit is lexical. It survives name removal and payoff rescaling but disengages when
Cooperate and Defect are replaced with neutral labels, and on 48
payoff-random games with neutral surfaces no model locks cooperative on
any dilemma or shows general equilibrium competence. In
mixed-model populations a single Nash-playing agent collapses
cooperation contagiously. What suppresses Nash play in large language
models is a word-triggered circuit rather than missing competence, and
it can be measured, bounded, and controlled.
\end{abstract}

large language models, activation steering, multi-agent systems,
opponent modeling

% -- 1. Introduction -----------------------------------------------------------
\section{Introduction}
\label{sec:intro}

Large language models have joined the short list of human artifacts too
complex to understand by derivation alone \citep{Broder2000}, so what
they do must be observed and why they do it must be measured from the
inside. This paper attempts both for a strategic behavior that comes
with a known normative benchmark.

Empirically, LLM agents often fail to converge to Nash equilibria in
repeated games \citep{Brookins2023, Akata2023, Fontana2025, JiaEtAl2025}.
When we run the grids ourselves, that aggregate sharpens into a split.
On the named Prisoner's Dilemma under direct prompting, three larger models, 32B to 72B across two families, lock at full
cooperation, $d = 2.00$, the metric's maximum, with zero variance
across replicates, while Llama-3-8B defects its way to near-Nash play. A failure this clean beside a success this clean is what a behavioral
account cannot explain and a mechanistic one must.

\looseness=-1 Opening the 8B, we find that the failure and the success are the same
computation resolved differently. Through roughly three quarters of the
network the residual stream favors the Nash action. A cooperative
override then surges, peaking at $0.95$ probability of cooperation at
layer 30, and the final layer corrects it away to $0.393$, which is why
this model plays Nash. When the same lens is applied to the larger
models, the surge turns out to be universal, and what varies is the
ending. The 70B corrects only to $0.821$ and cooperates, while Qwen-72B
reaches $0.589$ through a different internal architecture and cooperates
partially, so it is the size of the final correction, not the surge
itself, that rank-matches behavior across scale and architecture. The override is also a handle rather than just a picture. It is a
single archived direction in the residual stream, and clamping its
value at one position of one layer dials the decision monotonically,
Spearman $\rho = 1.000$ across the full range, from certain defection
to near-certain cooperation while generation stays fluent.
Re-extraction and random-direction controls establish what is special
about that particular direction. Finally, the effect has a
measured boundary. A prompt-surface ladder shows that the lock and the
internal signature are carried by the words Cooperate and Defect, and on
48 randomly generated games with neutral surfaces no model locks
cooperative on any dilemma and none reveals hidden equilibrium
competence, play landing within $0.1$ of an exact equilibrium in $42\%$
of games for Qwen-72B, $17\%$ for Qwen-32B, $8\%$ for Llama-8B, and
$0\%$ for Llama-70B.

\looseness=-1 The paper makes four contributions. The first is a behavioral and cross-play atlas
across four models, three reasoning modes, and all twelve ordered
pairings, whose headline is contagion. A single Nash-playing 8B
collapses any larger partner's cooperation, two large models sustain it
indefinitely, and equilibrium selection and payoff extraction are
anchored to seats in ways the distance metric cannot see. The second is a mechanistic account of the override, from trivially
decodable opponent information through the late surge and its
final-layer correction to its distribution across heads, established by
ablation with a verified positive control, plus the cross-scale
correction-magnitude correspondence. The third is causal control,
bidirectional steering, monotone clamping, and the controls separating
the direction's causal profile from the sample-sensitivity of its
extraction. The fourth is the boundary, where the ladder and the 48-game
experiment convert the generality question from an objection
into a measurement. 

The paper proceeds as follows. Section~\ref{sec:related} situates the
contribution in related work, Section~\ref{sec:setup} defines every
measured object, Sections~\ref{sec:behavioral}--\ref{sec:crossplay}
give the results, and Section~\ref{sec:discussion} states scope,
limitations, and deployment consequences.

% -- 2. Related Work -----------------------------------------------------------
\section{Related Work}
\label{sec:related}

\looseness=-1 Behaviorally, LLM deviation from equilibrium play is well documented.
GPT-4 cooperates in a human-like way, attributed to RLHF-induced
prosociality \citep{Brookins2023}, LLMs sustain
cooperation in repeated games given context \citep{Akata2023}, Llama
and GPT variants cooperate more than typical humans and forgive readily
\citep{Fontana2025}, and a 22-model evaluation finds chain-of-thought
helps but neither reasoning nor scale guarantees Nash approximation
\citep{JiaEtAl2025}. We treat this aggregate as our starting point. The contribution is the
internal account it leaves open, and our grids sharpen it into a clean
scale split on the named PD.

Mechanistically, prior work localizes computations to MLP layers
\citep{Meng2022} or attention heads \citep{Wang2023, McDougall2023}.
The override we study behaves instead as a distributed
residual-stream direction rather than a localized module, surviving
targeted head ablation under a verified positive control. The logit lens
\citep{Nostalgebraist2020, Belrose2023} observes late representational
shifts. We show one such shift moves with game history and under
intervention. The closest structural precedent probes an Othello-trained
transformer and recovers causal board-state representations \citep{LiEtAl2023}. Our
models are general-purpose, our games repeated with shared history, and
no game-specific training is involved.

Activation steering \citep{Turner2023, Zou2023} has reached economic
games twice so far. Demographic vectors shift Dictator-Game giving
\citep{MaEtAl2025}, and altruism persona vectors steered at one fixed
layer increase prosocial play across six games, with positive steering
stronger than negative \citep{SunZhang2026}. We differ in benchmarking
against exact equilibria, probing all layers, confirming causality by
projection-based clamping with re-extraction and random-direction
controls, and studying genuine 50-round two-agent interaction. The
positive-negative asymmetry they observe follows from the suppression
circuit we identify. \citet{GempEtAl2024} steer at the prompt level with
game-theoretic solvers rather than on activations.

% -- 3. Setup ------------------------------------------------------------------
\section{Experimental Setup and Measurement Definitions}
\label{sec:setup}

This section defines every object the paper measures, from the games
and the Nash-distance metric, through the models and the behavioral
protocol, to the mechanistic stimuli, the lens, the probe, and the
intervention readouts. All later sections use these definitions without
restating them.

\subsection{Games and the Nash Distance}
\label{sec:games_metric}

We use four two-player, two-action games covering the main equilibrium
structures, with payoffs small enough that every equilibrium is exact.
Section~\ref{sec:boundary} widens the atlas with 48 payoff-random
games under the same metric. In the \emph{Prisoner's Dilemma} (PD), Defect strictly
dominates. Mutual cooperation pays (3,3), mutual defection pays (1,1),
and unilateral defection pays the defector 5 against 0, so the unique
equilibrium is mutual defection while mutual cooperation is Pareto
superior. The \emph{Battle of the Sexes} (BoS) has two pure coordination
equilibria with opposed preferences, Opera favoring player A and
Football favoring player B, so it isolates which equilibrium a pair
selects. The \emph{Stag Hunt} (SH) has a payoff-dominant equilibrium
at Stag/Stag (4,4) and a risk-dominant one at Hare/Hare (3,3)
\citep{Harsanyi1988}.
\emph{Matching Pennies} (MP) is zero-sum with a unique mixed equilibrium
at 50/50, so equilibrium play requires sustained randomization.

To measure how far play is from equilibrium we track Nash distance.
After $t$ rounds let $\hat{\mu}_A^{(t)}$ and $\hat{\mu}_B^{(t)}$ be the
empirical mixed strategies of players A and B, and let $\mathcal{E}$ be
the game's set of Nash equilibria. Nash distance is the shortest
Euclidean distance from the empirical joint strategy to any equilibrium:
\begin{equation}
  d_{\text{Nash}}^{(t)} = \min_{(\sigma_A^*, \sigma_B^*) \in \mathcal{E}}
  \left\|
  \begin{pmatrix} \hat{\mu}_A^{(t)} - \sigma_A^* \\
                  \hat{\mu}_B^{(t)} - \sigma_B^* \end{pmatrix}
  \right\|_2
  \label{eq:nash_dist}
\end{equation}
A value of 0 means the pair is playing an equilibrium. In PD the maximum
of 2 means both players cooperated in every round. We use
$d_{\text{Nash}}^{(50)}$ as the summary statistic throughout.

Two caveats apply everywhere this metric appears. First,
$d_{\text{Nash}}$ is a distributional summary rather than a measure of
strategic coherence. A play sequence can match equilibrium frequencies
while remaining exploitable round to round, and
Section~\ref{sec:boundary} and Appendix~\ref{app:temp0} exhibit exactly
this pattern in the mixed-equilibrium games, so a low value should be
read as ``the empirical action frequencies resemble an equilibrium.''
Second, sampling at $\tau = 0.7$ is stochastic, so every self-play cell
is replicated across independent runs and reported as mean and standard
deviation (Appendix~\ref{app:ci}). The load-bearing cells are tight,
with each of the three larger models locking at full cooperation in PD
Direct mode ($d = 2.00$) at zero variance across replicates, while several
reasoning-mode cells are more variable and the corresponding claims are
scoped accordingly.

\subsection{Models, Decoding, and Behavioral Protocol}
\label{sec:protocol}

The experiments use four open-source instruction-tuned models,
Llama-3-8B-Instruct, Llama-3-70B-Instruct, Qwen2.5-32B-Instruct, and
Qwen2.5-72B-Instruct, spanning 8B to 72B parameters and two families so
that scale and architecture effects can be separated, all in fp16
through HuggingFace Transformers. The mechanistic experiments use the
8B exclusively, loaded through TransformerLens to modify activations at
each of its 32 layers, with lens checks on the two 70B-class models.
Behavioral runs sample at $\tau = 0.7$, and Appendix~\ref{app:temp0}
reports a greedy robustness grid in which the headline split reproduces
cell for cell.

Each game is played for 50 rounds of simultaneous moves. Each player's prompt shows the payoff matrix from that player's own
perspective, rows for its own actions, columns for the opponent's, its
own payoff first in each cell, so both players see their true
incentives. Play proceeds under three prompt structures. In
\emph{Direct} mode the model is asked only for an action name. In
\emph{Chain-of-Thought} (CoT) mode it reasons step by step before
committing \citep{Wei2022}. In \emph{Scratchpad} mode it reasons
privately before committing. Self-play covers 4 games $\times$ 3 modes
$\times$ 4 models. Cross-play adds all 12 ordered heterogeneous pairings
of the four models on the same grid.

One prompt-design detail matters for interpretation. The game
descriptions state the equilibrium in two of the four games, with PD
naming mutual defection and MP naming the 50/50 mixture, while BoS and
SH name none. Section~\ref{sec:boundary} tests this confound directly
with a prompt-ablation ladder on the PD structure, where removing the
equilibrium statement changes no cell while the action labels turn out
to carry both the behavioral and the mechanistic signature. A model that plays the
stated equilibrium may therefore be partly following the statement rather
than deriving it, and we flag this wherever it bears on a claim. The
exact prompts appear verbatim in Appendix~\ref{app:prompts}.

\subsection{Mechanistic Stimuli and Measurement Definitions}
\label{sec:measurement}

Every mechanistic and interventional quantity in this paper is evaluated
on one committed stimulus family, namely contrastive prompt sets
produced by a seeded generator (stimulus master seed 20260826) and
stored with
content hashes. Each set holds 50 prompt pairs for one game, built from
the player-A, chain-of-thought prompt of the behavioral protocol with a
synthetic history of 5 to 24 rounds. In the \emph{Nash-consistent}
prompt both players have repeatedly played the game's reference
equilibrium action, and in the \emph{deviating} prompt the opponent has
repeatedly played the other action. Reported curves average over the 50 Nash-consistent
prompts of a set unless stated otherwise.

The \emph{logit lens} reads the residual stream at the final token
position after each layer, applies the final normalization, projects
through the unembedding, and renormalizes over the two action tokens.
By construction the final-layer value equals the model's actual output
restricted to the two actions, verified programmatically in every run. The
\emph{opponent probe} is a logistic classifier on the same residual
representations (standardized features, five-fold balanced accuracy)
predicting which history the prompt contains. \emph{Interventions} use a
single archived unit vector $v_{\text{coop}}$, extracted once as a
layer-22 mean difference over contrastive activations and reused
unchanged in every experiment. Its extraction provenance and a
re-extraction robustness study appear in Section~\ref{sec:causal} and
Appendix~\ref{app:robustness}. Steering adds $\alpha\, v_{\text{coop}}$
to the residual stream at layers 20, 22, and 24 at all positions, while
clamping replaces the component of the layer-22 residual along
$v_{\text{coop}}$ at the final position with a fixed value $c$. The intervention readout is always the next-token distribution at the
decision position, renormalized over the two action tokens. The
quantitative results never rely on free-running generation, documented
separately in Appendix~\ref{app:intervention}.

% -- 4. Behavioral results -----------------------------------------------------
\section{Behavioral Results}
\label{sec:behavioral}

\begin{table}[t]
\centering
\caption{Self-play Nash distance $d^{(50)}$, mean $\pm$ SD over
replicates (15 for L-8B, 10 for the larger models), 50 rounds per run.
The BoS and MP cells use the role-perspective rendering of
Section~\ref{sec:protocol}; every such cell lies within one replicate SD of the earlier version's
value (Appendix~\ref{app:protocol}).}
\label{tab:behavioral}
\footnotesize
\begin{tabular}{llcccc}
\toprule
Game & Mode & L-8B & L-70B & Q-32B & Q-72B \\
\midrule
  PD & Direct & 0.04\,$\pm$\,0.01 & 2.00\,$\pm$\,0.00 & 2.00\,$\pm$\,0.00 & 2.00\,$\pm$\,0.00 \\
   & CoT & 0.04\,$\pm$\,0.05 & 0.00\,$\pm$\,0.01 & 0.44\,$\pm$\,0.75 & 0.29\,$\pm$\,0.29 \\
   & Scratchpad & 0.43\,$\pm$\,0.10 & 0.93\,$\pm$\,0.55 & 0.84\,$\pm$\,0.94 & 1.34\,$\pm$\,0.54 \\
\midrule
  BoS & Direct & 0.03\,$\pm$\,0.03 & 0.00\,$\pm$\,0.00 & 0.30\,$\pm$\,0.13 & 0.00\,$\pm$\,0.00 \\
   & CoT & 0.39\,$\pm$\,0.14 & 0.14\,$\pm$\,0.13 & 0.03\,$\pm$\,0.03 & 0.07\,$\pm$\,0.06 \\
   & Scratchpad & 0.30\,$\pm$\,0.10 & 0.14\,$\pm$\,0.07 & 0.08\,$\pm$\,0.06 & 0.21\,$\pm$\,0.17 \\
\midrule
  SH & Direct & 0.00\,$\pm$\,0.00 & 0.01\,$\pm$\,0.01 & 0.00\,$\pm$\,0.00 & 0.00\,$\pm$\,0.00 \\
   & CoT & 0.29\,$\pm$\,0.22 & 0.07\,$\pm$\,0.05 & 0.05\,$\pm$\,0.04 & 0.21\,$\pm$\,0.28 \\
   & Scratchpad & 0.54\,$\pm$\,0.12 & 0.09\,$\pm$\,0.05 & 0.02\,$\pm$\,0.02 & 0.30\,$\pm$\,0.27 \\
\midrule
  MP & Direct & 0.43\,$\pm$\,0.14 & 0.02\,$\pm$\,0.02 & 0.00\,$\pm$\,0.01 & 0.06\,$\pm$\,0.04 \\
   & CoT & 0.34\,$\pm$\,0.12 & 0.25\,$\pm$\,0.15 & 0.25\,$\pm$\,0.24 & 0.59\,$\pm$\,0.28 \\
   & Scratchpad & 0.17\,$\pm$\,0.12 & 0.12\,$\pm$\,0.03 & 0.11\,$\pm$\,0.09 & 0.21\,$\pm$\,0.14 \\
\bottomrule
\end{tabular}
\end{table}

Table~\ref{tab:behavioral} gives the behavioral picture across four
models and three reasoning conditions. The single most robust finding is a cooperative lock in the larger models.
In PD under Direct prompting, all three cooperate 100\% of the time,
$d = 2.00$ with zero variance across replicates, a complete failure to
play Nash that holds across both families and a more-than-twofold
spread in size. Llama-8B behaves in the opposite way, defecting
on almost every round without any reasoning ($d = 0.04$). In Direct mode the
failure to play Nash is therefore a property of the larger models rather
than a universal one, and the 8B, the model our mechanistic analysis opens
up, is the one model that does play Nash.
Section~\ref{sec:crossplay} shows that this single Nash-playing model makes
Nash play contagious in mixed pairings, and
Section~\ref{sec:boundary} shows that the lock itself is bound to the
game's surface presentation rather than to its payoff structure.

Chain-of-thought does not uniformly break the lock, and it does not worsen
the small model. The 8B stays near Nash under CoT ($d = 0.04$), Qwen-32B moves sharply
toward Nash on average ($0.44 \pm 0.75$, with large replicate
variance), Llama-70B reaches near-Nash ($d = 0.00$), and Qwen-72B is
variable ($0.29 \pm 0.29$). The clean lock is thus a Direct-mode
phenomenon of the larger models, consistent with a scale effect rather
than an architecture-specific one, though with only one sub-32B model
tested the scale reading remains suggestive. Scratchpad is the least
reproducible condition, with all three larger models highly variable in
PD (up to $1.34 \pm 0.54$), so we claim only that private reasoning
induces high-variance play, not any architecture ordering.

In the coordination games, Direct mode produces clean coordination
almost everywhere. Every model reaches a pure equilibrium in SH
($d \approx 0$) and three of four do in BoS ($d \leq 0.03$), with
Qwen-32B partially anti-coordinating ($0.30 \pm 0.13$). Which
equilibrium gets selected is now informative, because player B knows
that Football is its preferred equilibrium, and the coordinating pairs
nonetheless settle on Opera, player A's preferred outcome, in
essentially every Direct-mode round. Under the reasoning modes player B begins to
assert its own preference and coordination erodes (8B CoT
$0.39 \pm 0.14$). Section~\ref{sec:crossplay} gives the pairing-level analysis and the
first-listed confound.

\looseness=-1 MP is the hardest game. The 8B is the Direct-mode outlier
($0.43 \pm 0.14$), drifting toward a pure action instead of the required
50/50 mixture, while the larger models sit on the mixed frequencies
($d \leq 0.06$) by alternating actions at high rates. Here the metric caveat bites hardest, because alternation matches the
equilibrium frequencies while remaining exploitable, and
Appendix~\ref{app:temp0} shows greedy cases in which one player wins
every round at $d = 0.00$ against a correctly informed opponent that
never adapts. Low MP distances should be read accordingly.

Whether these deviations can be controlled is the question the
mechanistic analysis now takes up.

% -- 5. Mechanism --------------------------------------------------------------
\section{Mechanistic Analysis}
\label{sec:mechanism}

All mechanistic experiments open up the 32-layer Llama-3-8B-Instruct.
Layer numbers 0 through 31 refer to it unless stated otherwise, and
every measurement uses the committed stimuli and definitions of
Section~\ref{sec:measurement}. One scoping note carries over. The
mechanistic prompts do not apply the chat template the behavioral runs
use, so ``consistent with behavior'' statements are directional
correspondences rather than reproductions of the generation setup.
Section~\ref{sec:crossscale} extends the lens to Llama-3-70B and
Qwen2.5-72B.

\subsection{Opponent History Is Decodable but the Nash Action Is Not}
\label{sec:probe}

On the 100 committed PD prompts, a logistic probe on the
final-position residual stream decodes which history the prompt
contains with accuracy 1.000 at every layer.
Because the two conditions use homogeneous histories they are
perfectly separable by construction, so the result establishes that the
opponent's action is present everywhere in the residual stream, not
fine-grained tracking over mixed histories, which we do not test.

We deliberately report no Nash-action decodability number. Defining the
model's action as the argmax over the two action tokens at the decision
position, the 8B answers Defect on all 100 committed prompts, so the label
has no variance and a probe for it is undefined. This concerns the
argmax label rather than sampled play, since the mean probability mass
on Defect on these prompts is $0.607$, so the model favors the Nash action on every prompt
without placing all its mass there, itself consistent with the behavioral
finding that the 8B plays Nash. An earlier version of this paper reported
an apparent Nash-decodability signal. It was an artifact of pooling
trajectories from two code versions, and we rely on no such number.

\subsection{The Logit Lens Reveals the Override and Its Final-Layer Correction}
\label{sec:lens8b}

What a probe cannot show here, the lens can. On the committed PD set,
layers 0 through 23 assign majority probability to Defect. The picture
then inverts. $P(\text{Cooperate})$ surges through the late layers to a
peak of $0.952$ at layer 30 before the final layer corrects back toward
the Nash action, ending at $0.393$, which equals the model's true output
readout on these prompts. This late cooperative surge followed by a
final-layer correction (Figure~\ref{fig:crossscale}) is the computational
signature of the override in the model that behaviorally plays Nash. An
earlier version reported a peak near $0.80$ from an unseeded stimulus
draw. Under the seeded protocol two independent samples agree within
$0.012$ at every reported point, and all lens values in this paper come
from the committed sets (Appendix~\ref{app:protocol}).

\subsection{The Override Is Distributed Across Heads}
\label{sec:heads}

\begin{figure}[t]
  \centering
  \includegraphics[width=0.55\textwidth]{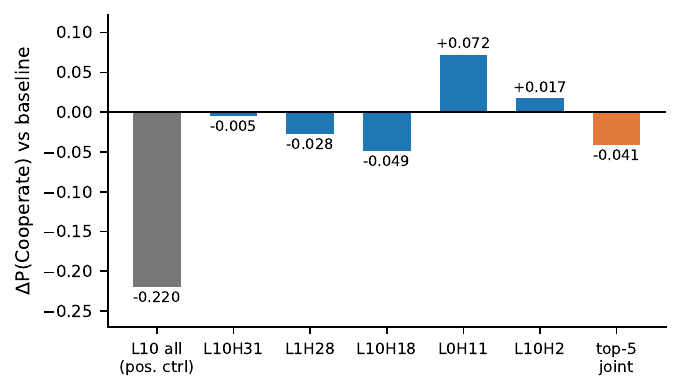}
  \caption{Zero-ablation deltas in $P(\text{Cooperate})$ relative to
  baseline on the committed PD prompts. The layer-10 whole-layer positive
  control confirms the ablation pathway is active. The top-5
  opponent-tracking heads, singly and jointly, produce only small shifts.}
  \label{fig:ablation}
\end{figure}

Can the override be attributed to specific attention heads? We score heads
by the attention they place on the opponent's past-action tokens and
zero-ablate the five top-scoring ones, individually and jointly, under
an ablation harness verified by a positive control, in which
zero-ablating all 32 heads of layer 10 shifts the two-token readout by
$0.220$, so the hooks demonstrably bite. Against that yardstick, no single top head moves the
readout by more than $0.072$ in absolute value, and ablating all five
jointly moves it by $0.041$ (Figure~\ref{fig:ablation}), an order of
magnitude smaller than the $0.6$ swing the residual-stream direction of
Section~\ref{sec:causal} produces. The override is a distributed
residual-stream effect. The lens shows it building gradually across the
late layers, and no head or combination among the top opponent-tracking heads we ablate
carries it, bounding those heads rather than all heads, which is why
steering on the full residual succeeds where head ablation fails.

\subsection{The Correction, Not the Surge, Tracks Behavior Across Scale and Architecture}
\label{sec:crossscale}

\looseness=-1 The same lens, formula, and stimuli on Llama-3-70B, an 80-layer model,
show the same signature, a late surge peaking at $0.970$ at layer 77. What differs is the ending. The 70B's
final readout eases only to $0.821$ whereas the 8B's collapses to
$0.393$, so both models compute the override and the size of the
final-layer correction is what separates them. Qwen2.5-72B shows a third pattern, reproduced here on the unified
protocol, with a mid-network collapse toward Defect (minimum
$P(\text{Cooperate}) = 0.131$ across layers 30--72) followed by an
abrupt late reversal spiking to $0.970$ at layer 77 and a partial
re-correction to a final readout of $0.589$.

\begin{figure}[t]
  \centering
  \includegraphics[width=0.814\textwidth]{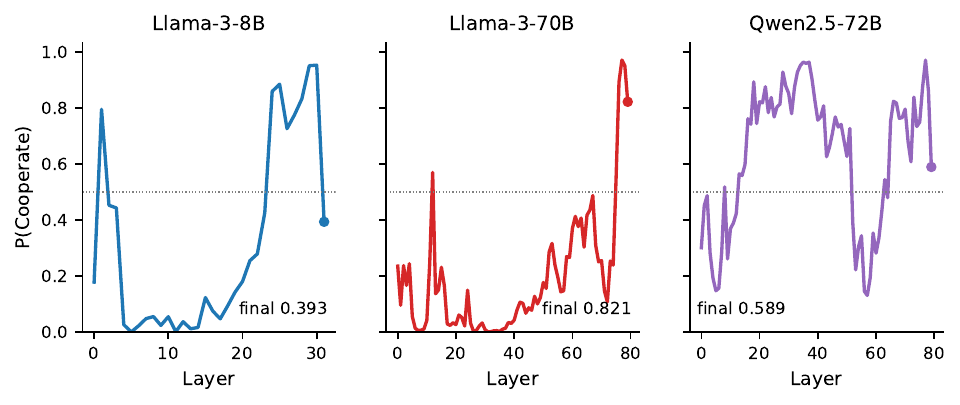}
  \caption{PD lens curves on the committed stimuli. All three models
  show the late cooperative surge. The final-layer correction magnitude
  rank-matches chain-of-thought behavior.}
  \label{fig:crossscale}
\end{figure}

Placing the three models side by side yields the cleanest
correspondence in the paper. The final lens readouts of $0.393$,
$0.821$, and $0.589$ rank-match the models' chain-of-thought PD
behavior, $d = 0.04$ near Nash, $0.00$ at full cooperation, and $0.29$
at partial cooperation respectively, and chain of thought is the mode
these stimuli are written in. The cooperative
surge is universal across the three models. The final correction
magnitude is what tracks who actually plays Nash
(Figure~\ref{fig:crossscale}). Every final value equals the model's true
output readout by construction, verified per run.

\begin{figure}[t]
  \centering
  \includegraphics[width=0.823\textwidth]{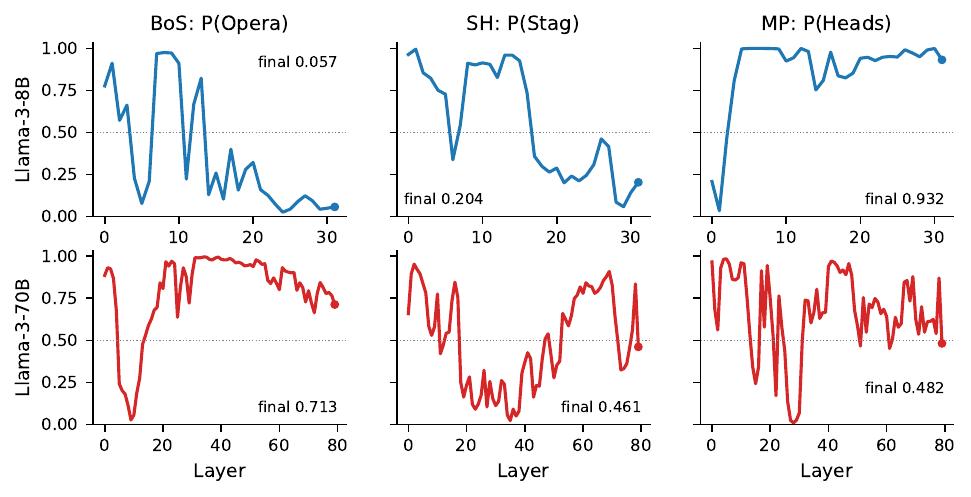}
  \caption{Lens curves for the three non-PD canonical games, committed
  stimuli; final readouts equal each model's true output distribution
  over the two actions.}
  \label{fig:crossgame_lens}
\end{figure}

Beyond PD, the cross-game panels (Figure~\ref{fig:crossgame_lens}) show
that each model's final readouts stay behaviorally coherent while the
canonical-game shapes differ by model. For the 70B, the MP readout ends
near chance ($0.482$), BoS ends Opera-leaning ($0.713$), matching its
Opera-heavy coordination, and SH ends mixed ($0.461$). The 8B's
non-PD canonical curves peak early rather than late and end at
$0.057$ (BoS), $0.204$ (SH), and $0.932$ (MP), values consonant with its
reasoning-mode play in those games (Table~\ref{tab:behavioral}). The
distinctively late surge with a large correction is, in the canonical
family, a PD phenomenon. Section~\ref{sec:boundary} shows that what
carries it is the words on the actions.

% -- 6. Causal control ---------------------------------------------------------
\section{Causal Control of the Override}
\label{sec:causal}

The override is not only visible; it is a handle. All interventions
use the archived unit vector $v_{\text{coop}}$ of
Section~\ref{sec:measurement} on the committed PD stimuli, with the
renormalized two-token readout at the decision position.

\subsection{Steering Produces a Bidirectional Dose-Response}
\label{sec:steering}

\begin{figure}[t]
  \centering
  \includegraphics[width=0.832\textwidth]{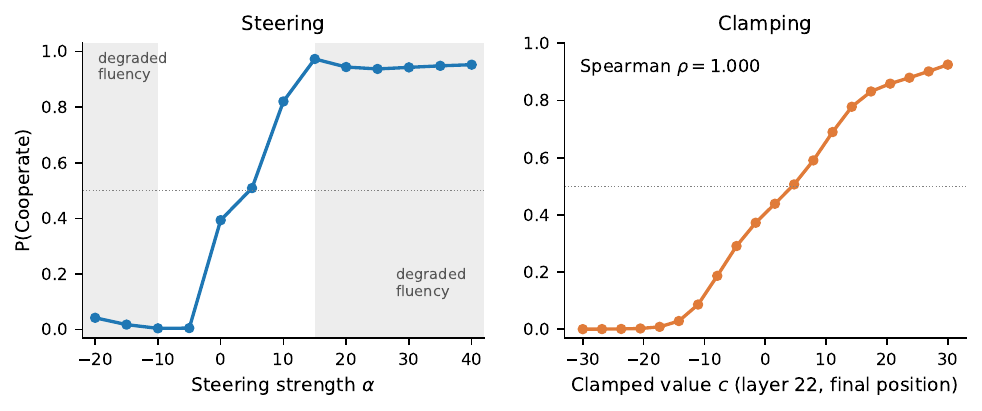}
  \caption{Causal control on the committed PD prompts. Left, the
  steering dose-response, with shaded bands marking the fluency-degraded
  regions. Right, concept clamping at layer 22, strictly monotone
  ($\rho = 1.000$) with fluent generation across the entire range.}
  \label{fig:steer_clamp}
\end{figure}

Adding $\alpha\, v_{\text{coop}}$ at layers 20, 22, and 24 moves the
decision smoothly in both directions
(Figure~\ref{fig:steer_clamp}). The unsteered baseline places $0.607$ on
Defect. At $\alpha = -5$ the model is at $0.995$ Defect, and the negative
side saturates near $0.996$ by $\alpha = -10$. On the positive side
cooperation rises through $0.509$ at $+5$ and $0.821$ at $+10$ to $0.974$
at $+15$, then holds a plateau near $0.95$ out to $+40$. A local
non-monotonicity at $+5$ reported in the previous version does not appear
under the seeded stimuli and was a prompt-sample artifact. Fluency
bounds how far the dial can be pushed. Greedy continuations remain
coherent at $\alpha = -5$, degrade visibly by $+10$, and collapse into
repetition at $\pm 20$ and beyond, so we treat the extremes as stress
tests of the readout rather than usable operating points, and
Appendix~\ref{app:intervention} reproduces the continuations verbatim.

\subsection{Clamping Is a Monotone Causal Dial}
\label{sec:clamping}

Steering adds a vector everywhere; clamping asks a sharper question,
replacing the residual component along $v_{\text{coop}}$ at layer 22,
final position only, with a fixed value $c$ and reading the decision. Across twenty clamp values from $-30$ to $+30$ the response is
strictly monotone, Spearman $\rho = 1.000$, running from
$P(\text{Cooperate}) = 0.000$ at $c = -30$ to $0.925$ at $c = +30$
(Figure~\ref{fig:steer_clamp}). The previously reported
$\rho = 0.725$ came from the unseeded stimuli, the same prompt-sample
artifact noted for steering above. Unlike the steering extremes, greedy
continuations stay fluent across the entire clamp range, including the
cooperative extreme, where the model continues ``Please make a decision,
and we,'' so the full dose-response lies inside the fluent regime. A single direction, set to a single value at a single position of
a single layer, dials the decision continuously from certain defection to
near-certain cooperation. This is the causal core of the paper.

\subsection{What Is Special About This Direction?}
\label{sec:robustness}

\looseness=-1 Is $v_{\text{coop}}$ the direction, or a direction? Re-running the
extraction on three fresh committed samples yields vectors that agree with each other almost perfectly (pairwise
cosine $\geq 0.978$) yet
are essentially orthogonal to the archived vector
($|\cos| \leq 0.016$), so extraction is stable but stably finds a
different axis on these stimuli. Under broad steering both axes move the
readout at sufficient strength, while three seeded random directions produce only erratic,
sign-inconsistent effects, so large-injection movement alone does not
certify a direction. The projection test does. Clamping works only if
the clamped component carries the decision, and it is fully effective
along the archived direction while nearly inert along the re-extracted
one ($P(\text{Cooperate}) \leq 0.51$ at $c = +30$ against $0.925$). The archived direction is thus privileged by the only
test that probes what the layer actually transmits, the extraction step
is the prompt-sensitive part, and Appendix~\ref{app:robustness} tabulates
all vectors, cosines, and control curves.

\subsection{Does the Direction Transfer Across Games?}
\label{sec:crossgame_steer}

\begin{figure}[t]
  \centering
  \includegraphics[width=0.99\textwidth]{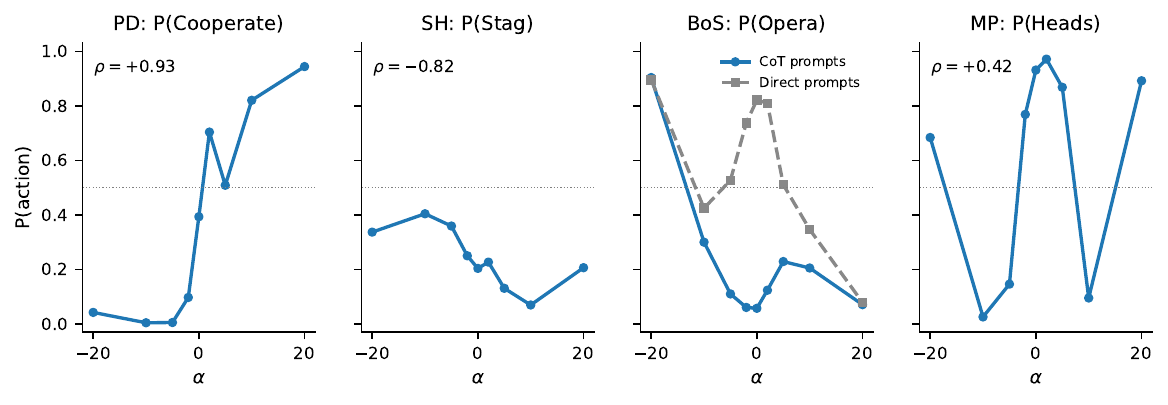}
  \caption{Cross-game steering with the unchanged archived vector, with
  graded control in PD and SH, an endpoint-only displacement in BoS
  (both prompt modes), and no control in MP.}
  \label{fig:crossgame_steer}
\end{figure}

Applied unchanged to the other canonical games on their committed
stimuli, the direction gives graded, directionally consistent control in
two of the three, in PD itself ($\rho = +0.93$ between $\alpha$ and
$P(\text{Cooperate})$) and in SH, where positive $\alpha$ moves play
toward Hare ($\rho = -0.82$ for Stag). Because SH uses different action
words and a different equilibrium structure, this transfer rules out the
concern that the direction merely encodes the Defect token. BoS marks the
boundary of transfer. No graded mid-range control appears in either
prompt mode ($|\rho| \leq 0.14$ for $|\alpha| \leq 10$), and what
survives is only a strong-$\alpha$ endpoint displacement toward Football
($P(\text{Opera}) = 0.08$ at $\alpha = +20$ in direct mode), inside
the fluency-degraded region. MP is
not controlled ($\rho = +0.42$, not significant against the sweep's
noise). The transfer is therefore real but partial, and we report it as
such (Figure~\ref{fig:crossgame_steer}).

% -- 7. Boundary conditions ----------------------------------------------------
\section{Boundary Conditions}
\label{sec:boundary}

The results so far concern four named games that are abundantly present in
training data. This section determines what actually carries the effects,
using two instruments, a prompt-surface ablation ladder that varies the
surface of the PD while holding its structure fixed, and 48 randomly
generated games with neutral surfaces that hold nothing fixed at all.

\subsection{The Prompt-Surface Ladder Shows the Lock Is Lexical}
\label{sec:ladder}

The ladder runs five conditions in self-play Direct mode, one greedy
and three to five sampled runs per model per condition. It begins from
the named game, then removes the equilibrium-stating hint, then moves
to a neutral task framing with the canonical payoffs, then rescales
the
payoffs affinely, and finally disguises the game fully by replacing the
labels Cooperate and Defect with Circle and Square.
Table~\ref{tab:ladder_behavior} gives Player A's cooperation rate in each.

\begin{table}[t]
\centering
\small
\caption{Cooperation rate of Player A at $\tau = 0.7$, mean $\pm$ SD over
independent runs, with the greedy run in parentheses. Named-condition values are Table~\ref{tab:behavioral}'s; rescaled payoffs
are the affine $2x+1$ of the canonical payoffs, preserving equilibria
exactly.}
\label{tab:ladder_behavior}
\footnotesize
\setlength{\tabcolsep}{3.5pt}
\begin{tabular}{lcccc}
\toprule
Condition & L-8B & L-70B & Q-32B & Q-72B \\
\midrule
Named             & $0.02$ & $1.00$ & $1.00$ & $1.00$ \\
Hint-ablated      & $0.03 \pm 0.01$ ($0.02$) & $1.00 \pm 0.00$ ($1.00$) & $1.00 \pm 0.00$ ($1.00$) & $0.91 \pm 0.13$ ($1.00$) \\
Unnamed           & $0.51 \pm 0.39$ ($1.00$) & $1.00 \pm 0.00$ ($1.00$) & $1.00 \pm 0.00$ ($1.00$) & $1.00 \pm 0.00$ ($1.00$) \\
Unnamed, rescaled & $1.00 \pm 0.00$ ($1.00$) & $1.00 \pm 0.00$ ($1.00$) & $1.00 \pm 0.00$ ($1.00$) & $0.89 \pm 0.16$ ($1.00$) \\
Fully disguised   & $0.72 \pm 0.32$ ($1.00$) & $0.52 \pm 0.13$ ($0.74$) & $0.36 \pm 0.29$ ($0.08$) & $0.47 \pm 0.21$ ($1.00$) \\
\bottomrule
\end{tabular}
\end{table}

Three findings follow, one per ablated factor. First, the equilibrium
statement is behaviorally inert. Removing only the hint changes no cell,
so the prompt-leakage confound of Section~\ref{sec:protocol} does not
drive the results. Second, the action labels carry the lock. Wherever the
labels read Cooperate and Defect the larger models cooperate at or near
$100\%$ regardless of framing, hint, or payoff values, and the
single-factor swap to Circle and Square dissolves the lock into noisy
mixed play in all four models. The lens shows the same gate internally:
on the committed stimuli, all four Cooperate/Defect conditions reproduce
the 8B's late surge and correction (peaks $0.932$ to $0.952$ at layers 29
to 30, finals $0.359$ to $0.393$), while the disguised condition has no
late surge at all, peaking at layer 8 ($0.993$) and declining
monotonically to $0.150$. Third, the 8B's near-Nash play is keyed to the
framing rather than the matrix. On the identical payoffs presented as a
neutral task it flips from $2\%$ to $100\%$ cooperation under greedy decoding
while its internal lens profile is unchanged, so its defection policy
tracks recognition of the named game rather than a computation over
payoffs.

\subsection{Random Neutral Games Show Neither Lock nor Equilibrium Competence}
\label{sec:randomgames}

The ladder varies one structure; the random games vary all of them. We
generate 48 two-player $2 \times 2$ games with integer payoffs, stratified over the five generic structural classes, with six social
dilemmas, six other double-dominance games, twelve single-dominance
games, twelve with two pure equilibria, and twelve with only a mixed
equilibrium. Every equilibrium is computed exactly and independently
cross-checked, the surface is the ladder's neutral template with Circle
and Square randomly assigned per game, and every cell runs at both
temperatures with three sampled replicates (Appendix~\ref{app:randomgames} lists the
roster and per-game results).

At the paper's temperature, no model locks cooperative on a single
random dilemma. Table~\ref{tab:random_dilemmas} shows the six dilemmas.
The three models that cooperate at $100\%$ with zero variance on the
named PD cooperate at $0$ to $39\%$ here, and Qwen-72B, the strongest
named-game cooperator, sits at $0$ to $1\%$ on five of the six while
playing the equilibrium with $90\%$ dominant-action rates. The occasional greedy locks are knife-edge. Qwen-72B locks on one
dilemma under greedy decoding ($0.98$ cooperation) and the lock
vanishes under sampling ($0.01$), the 8B's second greedy lock softens
from $0.98$ to $0.31$, and the named-game lock is invariant across both
decoding regimes. Each model that ever locks does so on different,
payoff-arbitrary games, idiosyncrasy rather than structure.

\begin{table}[t]
\centering
\small
\caption{The six random social dilemmas at $\tau = 0.7$, showing the
cooperation rate with the Nash distance in parentheses, replicate means. rg01 and rg04 are the
two games on which the 8B locks cooperative under greedy decoding.}
\label{tab:random_dilemmas}
\footnotesize
\begin{tabular}{lcccc}
\toprule
Game & L-8B & L-70B & Q-32B & Q-72B \\
\midrule
rg01 & 0.95 (1.92) & 0.39 (1.12) & 0.07 (0.43) & 0.00 (0.08) \\
rg02 & 0.02 (0.19) & 0.25 (0.79) & 0.05 (0.57) & 0.01 (0.15) \\
rg03 & 0.01 (0.19) & 0.08 (0.51) & 0.08 (0.62) & 0.00 (0.41) \\
rg04 & 0.31 (1.40) & 0.35 (1.05) & 0.11 (0.69) & 0.00 (0.20) \\
rg05 & 0.00 (0.98) & 0.07 (0.47) & 0.05 (0.43) & 0.00 (0.17) \\
rg06 & 0.01 (0.36) & 0.09 (0.49) & 0.04 (0.41) & 0.01 (0.45) \\
\bottomrule
\end{tabular}
\end{table}

\looseness=-1 Removing the surface does not reveal hidden equilibrium
competence either. At
$\tau = 0.7$, empirical play lies within $0.1$ of an exact equilibrium in $42\%$ of the 48 randomly
generated games for Qwen-72B, in $17\%$ for Qwen-32B, in $8\%$ for
Llama-8B, and in $0\%$ for Llama-70B, with the greedy figures at $54$,
$27$, $46$, and $4\%$. Success concentrates where structure is easiest, Qwen-72B solving
every double-dominance game from the first round under greedy, and
vanishes where it is hardest. In the twelve games with only a mixed
equilibrium no model at either temperature comes near it, and under greedy decoding the 8B freezes on a fixed action profile
from the second round in five of the twelve, one player then losing
every remaining round, by up to $391$ payoff points, against a
correctly
informed, never-adapting opponent. The named-game scale story thus inverts on neutral ground, an uneven
ordering with the 70B last, and the metric caveat of
Section~\ref{sec:games_metric} does double duty, since low distance never
meant strategic soundness and here high distance accompanies outright
strategic incoherence.

\subsection{The Surge Needs the Words}
\label{sec:random_lens}

\begin{figure}[t]
  \centering
  \includegraphics[width=0.832\textwidth]{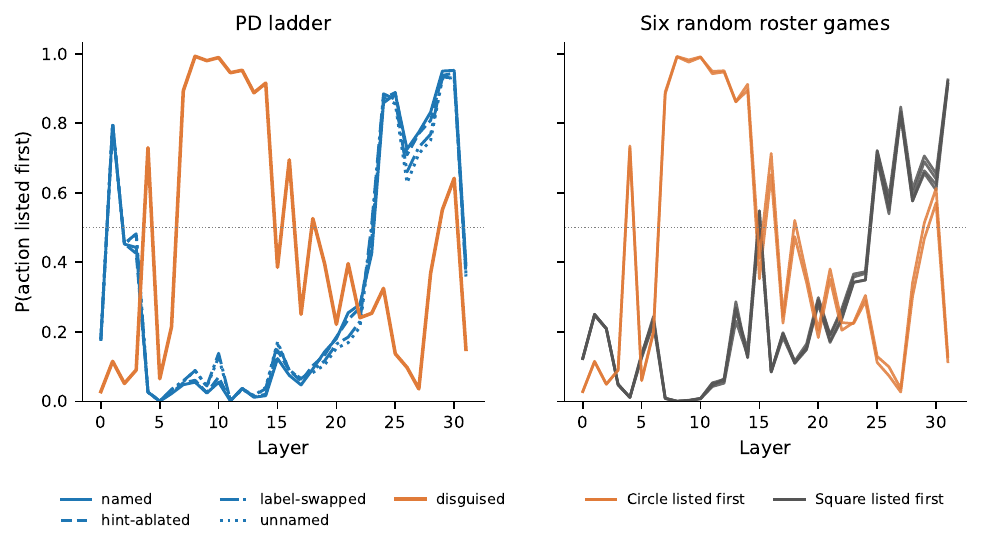}
  \caption{Lens curves for the four Cooperate/Defect PD conditions, the
  disguised condition, and the six random roster games; committed
  stimuli throughout.}
  \label{fig:ladder_lens}
\end{figure}

Six roster games, one per stratum plus both dilemmas on which the 8B
locks under greedy decoding, were added to the committed lens suite.
None shows the late cooperative surge, including the games where the
8B behaviorally locks cooperative. The six curves fall into
exactly two shapes, and the split follows the label listed first rather than the
game's structure or the model's behavior. Circle-first sets show an early
transient at layer 8 and then decline monotonically, identical in shape
to the disguised ladder condition, while Square-first sets rise late and
stay. Read jointly, every neutral-label set ends with the final readout
simply tracking the repeated history action ($0.85$ to $0.93$), whereas
every set carrying the words Cooperate and Defect holds its final readout
near $0.37$ cooperate against fifty rounds of opponent defection, after
the distinctive late surge. The override's trigger is lexical, and this rests not only on the
ladder's single disguised game but on six payoff-random structures
(Figure~\ref{fig:ladder_lens}).

Together the ladder and the random games set the paper's scope. The
cooperative lock, the override, and the 8B's Nash play are properties of
the games as presented, engaged by the cooperation vocabulary, and
neither the lock nor general equilibrium play survives on neutral
payoff-random ground.

% -- 8. Cross-play -------------------------------------------------------------
\section{Cross-Play as Application}
\label{sec:crossplay}

Self-play characterizes each model against a copy of itself.
Deployments mix models. All 12 ordered pairings played all four games
in all three modes, 144 cells at 10 replicates, and the mechanism of
Sections~\ref{sec:mechanism}--\ref{sec:causal} turns out to have a
population-level consequence that self-play cannot show.

\begin{table}[t]
\centering
\caption{Cross-play Nash distance $d^{(50)}$ in Direct mode, mean
$\pm$ SD over 10 replicates, all 12 ordered pairings; BoS and MP columns
under the role-perspective rendering.}
\label{tab:crossplay}
\footnotesize
\setlength{\tabcolsep}{4pt}
\begin{tabular}{llcccc}
\toprule
Agent A & Agent B & PD & BoS & SH & MP \\
\midrule
\multicolumn{6}{l}{\emph{Mixed pairings (one 8B model present)}}\\
  L-70B & L-8B & 0.07\,$\pm$\,0.02 & 0.03\,$\pm$\,0.05 & 0.00\,$\pm$\,0.00 & 0.29\,$\pm$\,0.17 \\
  L-8B & L-70B & 0.07\,$\pm$\,0.02 & 0.00\,$\pm$\,0.00 & 0.07\,$\pm$\,0.07 & 0.31\,$\pm$\,0.15 \\
  L-8B & Q-32B & 0.09\,$\pm$\,0.04 & 0.03\,$\pm$\,0.01 & 0.00\,$\pm$\,0.00 & 0.38\,$\pm$\,0.13 \\
  L-8B & Q-72B & 0.07\,$\pm$\,0.02 & 0.05\,$\pm$\,0.10 & 0.00\,$\pm$\,0.00 & 0.32\,$\pm$\,0.14 \\
  Q-32B & L-8B & 0.09\,$\pm$\,0.04 & 0.04\,$\pm$\,0.03 & 0.00\,$\pm$\,0.00 & 0.33\,$\pm$\,0.20 \\
  Q-72B & L-8B & 0.07\,$\pm$\,0.02 & 0.02\,$\pm$\,0.02 & 0.00\,$\pm$\,0.00 & 0.35\,$\pm$\,0.15 \\
\midrule
\multicolumn{6}{l}{\emph{Both models $\geq$ 32B}}\\
  L-70B & Q-32B & 2.00\,$\pm$\,0.00 & 0.18\,$\pm$\,0.10 & 0.00\,$\pm$\,0.00 & 0.01\,$\pm$\,0.01 \\
  L-70B & Q-72B & 2.00\,$\pm$\,0.00 & 0.03\,$\pm$\,0.05 & 0.00\,$\pm$\,0.00 & 0.02\,$\pm$\,0.02 \\
  Q-32B & L-70B & 2.00\,$\pm$\,0.00 & 0.00\,$\pm$\,0.00 & 0.01\,$\pm$\,0.01 & 0.02\,$\pm$\,0.02 \\
  Q-32B & Q-72B & 2.00\,$\pm$\,0.00 & 0.01\,$\pm$\,0.02 & 0.00\,$\pm$\,0.00 & 0.03\,$\pm$\,0.02 \\
  Q-72B & L-70B & 2.00\,$\pm$\,0.00 & 0.00\,$\pm$\,0.00 & 0.01\,$\pm$\,0.01 & 0.04\,$\pm$\,0.03 \\
  Q-72B & Q-32B & 2.00\,$\pm$\,0.00 & 0.07\,$\pm$\,0.05 & 0.00\,$\pm$\,0.00 & 0.03\,$\pm$\,0.03 \\
\bottomrule
\end{tabular}
\end{table}

The first effect is the defection unlock. The 8B plays close to Nash in
PD, and its presence makes Nash play contagious. Every mixed pairing
collapses to mutual defection ($d \approx 0.08$), although the larger
partner cooperates at $100\%$ in self-play and in every all-large
pairing, and the mechanism requires no strategic depth, since the 8B
defects, the larger model best-responds to a defector by defecting, and
mutual defection follows. Between large models the opposite equilibrium
of the population holds. Every pairing of two models at or above 32B
sits at $d = 2.00$ with zero variance, each sustaining the other's
cooperative prior indefinitely. Chain-of-thought weakens both effects, with substantial replicate
variance, dissolving neither cleanly.

\looseness=-1 The BoS and MP data add a finding the distance metric conceals. \looseness=-1 In BoS, eleven of the twelve Direct pairings coordinate mostly or
entirely on Opera, seat A's preferred equilibrium, whichever model occupies which seat, although seat B is correctly
informed that Football pays it more (the partial exception, Q-72B
seated A against Q-32B). Selection is anchored to the seat rather than the model, with
the caveat that Opera is also the first-listed action, so seat
preference and first-label bias cannot be separated within this one
game. In MP, the all-large pairings hold $d \leq 0.04$ by
synchronized high-frequency alternation, yet the payoffs underneath are
lopsided, with seat A skimming up to $+64$ points over 50 rounds in
Direct mode, and under chain-of-thought the edge flips to seat B in
eleven of twelve pairings. The pairings containing the 8B instead break the alternation
entirely ($d \approx 0.3$), consistent with its self-play difficulty
randomizing.

At the level of $d$, role order is symmetric in the clean Direct
condition, as the earlier version of this paper reported. The
round-level data shows why that summary was incomplete, since
order-symmetric distances coexist with seat-anchored selection and
payoff extraction. Which equilibrium a pair lands on, and
who profits inside a frequency-symmetric dance, are seat effects $d$ cannot see.

% -- 9. Discussion and conclusion ----------------------------------------------
\section{Discussion, Limitations, and Conclusion}
\label{sec:discussion}

\subsection{What the Mechanism Is}

Through roughly three quarters of the 8B's depth the intermediate
readouts lean toward the Nash action, a distributed override then surges, peaking near $0.95$,
and the final layer settles the contest, collapsing to $0.393$ in the
8B, easing to $0.821$ in the 70B, and re-correcting to $0.589$ in
Qwen-72B. It is distributed enough that ablating the top opponent-tracking heads
barely moves it, against a verified positive control, yet concentrated
enough that one direction clamps the decision monotonically. The base-model comparison (Appendix~\ref{app:base_model}) shows the
override precedes instruction tuning, which modulates rather than creates it. The steering asymmetry of \citet{SunZhang2026}, strong toward altruism and weak away from it, follows as well, since pushing the
direction up amplifies an active process while pushing it down can only
suppress one.

\subsection{Scope, Limitations, and Conclusion}

\looseness=-1 The lock, the override signature, and the 8B's Nash play are
properties of the games as presented, engaged by the cooperation
vocabulary. None survives on payoff-random ground, and no general equilibrium
competence appears there either.
Six limitations travel with them. Causal interventions are established for the 8B only, with lens
corroboration at scale. The mechanistic prompts do not apply the chat
template the behavioral runs use, so correspondences are directional.
Cross-game transfer is partial, graded in PD and SH, endpoint-only in
BoS, absent in MP. Re-extraction on fresh stimuli stably finds a
different, clamp-inert axis, so extraction, not the causal profile, is prompt-sensitive, and the
archived vector's privilege rests on the projection test. The seat effects in BoS are confounded
with first-listed-label bias. The random-games evidence covers $2 \times 2$ structures at 50
rounds. Larger action spaces and longer horizons are untested. In mixed
populations a single Nash-playing agent collapses cooperation
contagiously, two large models sustain it, and low Nash distance can coexist with
systematic exploitation, so equilibrium-frequency metrics alone are not
safety evidence.

\looseness=-1 On the named Prisoner's Dilemma, larger models fail to play Nash
because a cooperative override wins the final-layer contest. In the model that plays Nash the surge is corrected away, its size
tracking behavior across scale and architecture. The override is a single causally controllable direction. Clamping it
dials the decision monotonically with fluent generation. It is lexical,
carried by the words Cooperate and Defect, inert under neutral labels,
absent across 48 payoff-random games showing neither lock nor
equilibrium play. What suppresses Nash play is a word-triggered circuit rather than
missing competence, and it can be measured,
bounded, and controlled.

\section*{Broader Impact Statement}

This paper identifies and causally controls a mechanism suppressing
equilibrium play, and the control is dual-use. Clamping the override
toward Nash makes an agent more exploitative in cooperative settings,
in negotiation, auctions, and common-pool interactions, where the
cooperative side concedes value. Cross-play adds a population-level
warning, since a single Nash-playing agent collapses cooperation
contagiously, and low Nash distance can coexist with systematic
exploitation, so equilibrium-frequency metrics alone are not safety
evidence.

\bibliographystyle{plainnat}
\bibliography{references}

\appendix

\section{Prompt Templates}
\label{app:prompts}

The behavioral experiments use the following prompt structure, applied
through the model's chat template. The mechanistic analyses use the same
textual structure as a plain string, without the chat template, with
synthetic constructed histories (Section~\ref{sec:measurement}). Each
prompt is built from a fixed system prompt, a game description with
payoffs rendered from the receiving player's own perspective, the
accumulated round history, and a decision request whose final line is the
only thing the reasoning mode changes.

\subsubsection*{System Prompt}
\begin{quote}\small
You are a rational agent playing a strategic game for real payoffs.
Your goal is to maximize your own cumulative payoff over many rounds.
Think carefully about your opponent's likely strategy based on past play.
\end{quote}

\subsubsection*{Game Descriptions}

\textbf{Prisoner's Dilemma} (exact frozen prompt; the description states
the Nash equilibrium, the confound discussed in
Section~\ref{sec:protocol}):
\begin{quote}\small
GAME: Prisoner's Dilemma\\
Classic social dilemma. Nash: both defect, but mutual cooperation is Pareto optimal.\\
You are Player A. Payoffs shown as (Your payoff, Opponent payoff).\\
Actions: Cooperate / Defect\\
Payoff matrix:\\
\hspace*{1em}[Opp: Cooperate] | [Opp: Defect]\\
\hspace*{1em}Cooperate: (3,3) | (0,5)\\
\hspace*{1em}Defect: (5,0) | (1,1)
\end{quote}

\textbf{Battle of the Sexes} (summary form; the frozen prompt names no
equilibrium):
\begin{quote}\small
You are playing a game called Battle of the Sexes.
Each round you choose Opera or Football.
Payoffs: both choose Opera (2,1), both choose Football (1,2), mismatch (0,0).
\end{quote}

\textbf{Stag Hunt} (summary form; no equilibrium named):
\begin{quote}\small
You are playing a game called Stag Hunt.
Each round you choose Stag or Hare.
Payoffs: both hunt Stag (4,4), both hunt Hare (3,3),
one hunts Stag alone gets 0, Hare alone gets 3.
\end{quote}

\textbf{Matching Pennies} (exact frozen prompt; like PD, the description
states the equilibrium):
\begin{quote}\small
GAME: Matching Pennies\\
Zero-sum game. Unique mixed Nash at 50/50. Tests randomization and exploitation detection.\\
You are Player A. Payoffs shown as (Your payoff, Opponent payoff).\\
Actions: Heads / Tails\\
Payoff matrix:\\
\hspace*{1em}[Opp: Heads] | [Opp: Tails]\\
\hspace*{1em}Heads: (1,-1) | (-1,1)\\
\hspace*{1em}Tails: (-1,1) | (1,-1)
\end{quote}

\textbf{Random roster games} (exact frozen template of
Section~\ref{sec:randomgames}, shown for rg01 as rendered for Player A;
every roster game uses this neutral surface with its own payoffs and its
own randomized assignment of Circle and Square to the action indices):
\begin{quote}\small
GAME: Two-Player Decision Task\\
Repeated two-player task. Each round, both players choose one action
simultaneously; payoffs for each pair of choices are shown in the matrix
below.\\
You are Player A. Payoffs shown as (Your payoff, Opponent payoff).\\
Actions: Square / Circle\\
Payoff matrix:\\
\hspace*{1em}[Opp: Square] | [Opp: Circle]\\
\hspace*{1em}Square: (5,6) | (7,2)\\
\hspace*{1em}Circle: (4,8) | (6,7)
\end{quote}

\subsubsection*{History Block}
\begin{quote}\small
Game history so far: \\
\ \ Round 1: You played [Action], Opponent played [Action]. \\
\ \ Round 2: You played [Action], Opponent played [Action]. \\
\ \ $\ldots$
\end{quote}

\subsubsection*{Decision Instruction by Reasoning Mode}
\textbf{Direct:}
\begin{quote}\small
Round $N$: What do you choose? Answer with exactly one word: [Action1] or [Action2].
\end{quote}
\textbf{Chain-of-Thought:}
\begin{quote}\small
Round $N$: Think step by step about what to do, then state your final answer as exactly one word: [Action1] or [Action2].
\end{quote}
\textbf{Scratchpad:}
\begin{quote}\small
Round $N$: You may reason privately before deciding. Your reasoning will not be shown to your opponent. Then state your final answer as exactly one word: [Action1] or [Action2].
\end{quote}

\section{The Cooperative Override Is Present in the Base Model}
\label{app:base_model}

Does the override require instruction tuning? Throughout the paper,
``the 8B'' denotes Llama-3-8B-Instruct. Here we ran the logit lens on
its base counterpart, Meta-Llama-3-8B (HuggingFace identifier
\texttt{meta-llama/Meta-Llama-3-8B}), the pretrained model sharing
identical architecture but never instruction tuned. If the override were installed by alignment fine-tuning, it
should be absent or weaker in the base model. It is neither.

\begin{figure}[t]
  \centering
  \includegraphics[width=0.82\textwidth]{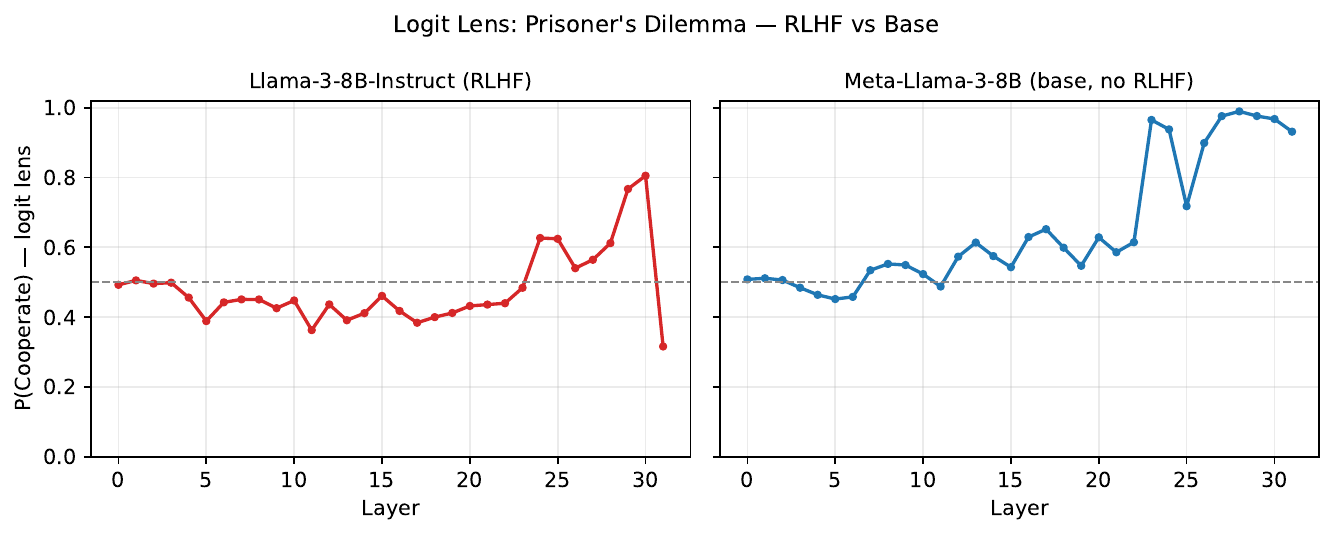}
  \caption{Logit lens comparison of Llama-3-8B-Instruct (left) and the
  base Meta-Llama-3-8B (right), collected under the earlier version's
lens protocol (Appendix~\ref{app:protocol}). The override peaks
  near $0.99$ in the base model and never corrects, whereas the instruct
  model corrects in the final layer. The override precedes instruction
  tuning.}
  \label{fig:base_logitlens}
\end{figure}

The override is present in the base model and is in fact stronger
(Figure~\ref{fig:base_logitlens}). In the base model the cooperative
tendency builds earlier, already exceeding 0.60 by layer 13, peaks near
0.99 at layers 27--28, and never corrects, so the base model cooperates
completely in its final readout, whereas the instruct model shows the
final-layer correction toward the Nash action documented in the main
text. Both curves in this appendix were produced under the earlier,
unseeded stimulus protocol; the instruct-side signature under the seeded
protocol appears in Section~\ref{sec:lens8b} (peak $0.952$, final
$0.393$), and the qualitative contrast between the two models is
unaffected by the protocol.

The scoped conclusion is that the cooperative prior precedes instruction
tuning and is already fully formed in the pretrained model, with
instruction tuning restraining rather than creating it. A natural
interpretation, which we offer as interpretation and not as a measured
result, is that the bias is absorbed from pretraining on human-generated
text, whose tone is predominantly cooperative.

\section{Reproducibility Across Replicates}
\label{app:ci}

To assess reproducibility we ran every self-play cell across multiple
independent replicates, 15 for Llama-3-8B and 10 for the larger models
at 50 rounds each, and report the distribution of the final Nash
distance $d^{(50)}$. Because the rounds of a repeated game are dependent and per-round
behavior drifts over the course of a game, the appropriate unit of
variation is the independent run rather than the round, and the
replicate spread below is that measurement.

Two patterns are worth stating. First, the load-bearing results are
highly reproducible. The cooperative lock of the three larger models in
PD Direct mode sits at $d = 2.00$ with zero variance across all
replicates, and the near-Nash Direct-mode cells have standard deviations
at or near zero. Second, several CoT and Scratchpad cells are genuinely
variable, with the largest spreads for the larger models in PD under
reasoning (Qwen-32B CoT $0.44 \pm 0.75$, Qwen-32B Scratchpad
$0.84 \pm 0.94$, Qwen-72B Scratchpad $1.34 \pm 0.54$, Llama-70B
Scratchpad $0.93 \pm 0.55$), and claims that depend on those cells are
scoped accordingly in the main text.
Table~\ref{tab:reproducibility} gives the full per-cell spread for
Llama-3-8B, the model used for the mechanistic analysis; the complete
replicate grids for all four models are available from the authors upon
reasonable request.

\begin{table}[h]
\centering
\caption{Reproducibility of the Llama-3-8B self-play Nash distance across
15 independent replicates at 50 rounds per run, mean and standard
deviation of $d^{(50)}$. Direct-mode and near-Nash cells are highly
reproducible, while several CoT and Scratchpad cells are variable, and
the corresponding claims are scoped accordingly.}
\label{tab:reproducibility}
\begin{tabular}{llccc}
\toprule
Game & Mode & Mean $d^{(50)}$ & SD & Replicates \\
\midrule
  PD & Direct & 0.045 & 0.010 & 15 \\
   & CoT & 0.040 & 0.053 & 15 \\
   & Scratchpad & 0.432 & 0.103 & 15 \\
\midrule
  BoS & Direct & 0.030 & 0.025 & 15 \\
   & CoT & 0.388 & 0.143 & 15 \\
   & Scratchpad & 0.300 & 0.103 & 15 \\
\midrule
  SH & Direct & 0.000 & 0.000 & 15 \\
   & CoT & 0.292 & 0.222 & 15 \\
   & Scratchpad & 0.541 & 0.120 & 15 \\
\midrule
  MP & Direct & 0.432 & 0.135 & 15 \\
   & CoT & 0.336 & 0.118 & 15 \\
   & Scratchpad & 0.168 & 0.117 & 15 \\
\bottomrule
\end{tabular}
\end{table}

\begin{figure}[t]
  \centering
  \includegraphics[width=0.7\textwidth]{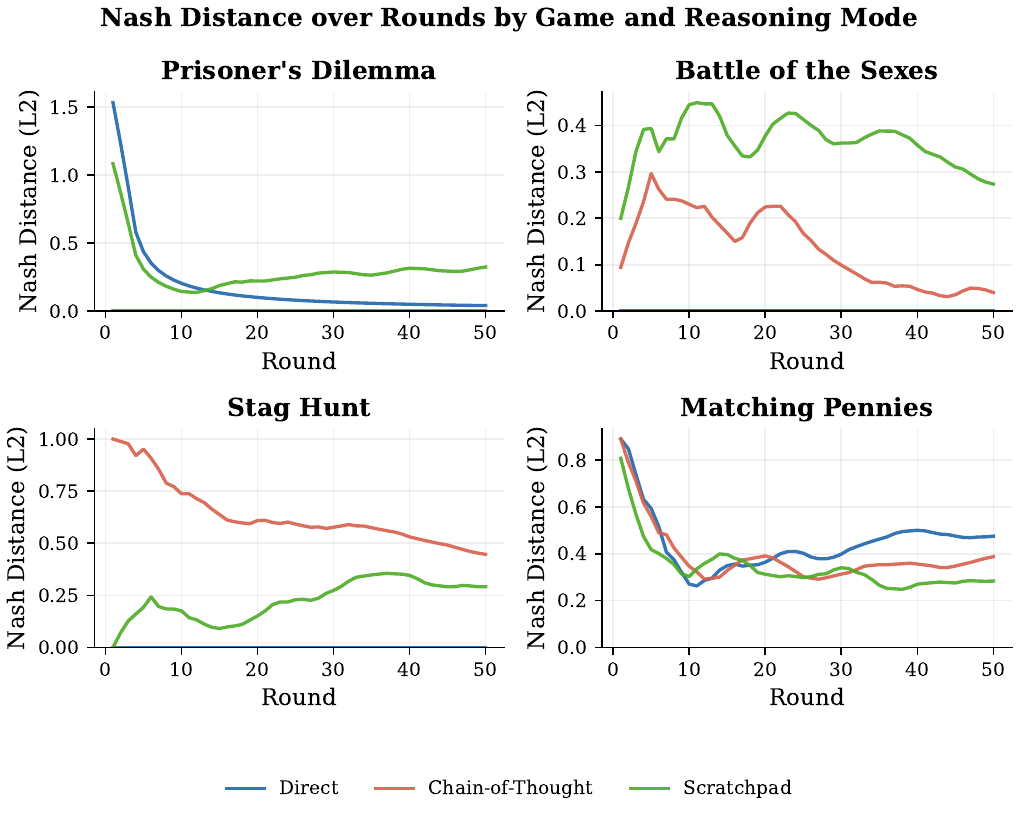}
  \caption{Nash distance over 50 rounds, Llama-3-8B self-play, one
  representative seed per reasoning mode. Under Direct prompting the 8B
  plays close to Nash in PD and SH, CoT and Scratchpad sit farther from
  equilibrium in the coordination games, and in MP both reasoning modes
  are closer than Direct.}
  \label{fig:convergence_appendix}
\end{figure}

Figure~\ref{fig:convergence_appendix} shows the round-by-round Nash
distance for one representative Llama-3-8B replicate per mode, the
within-run trajectory whose endpoint the tables above summarize.

\section{Robustness to Decoding Temperature (Greedy, $\tau = 0$)}
\label{app:temp0}

All main results use sampled decoding at $\tau = 0.7$ with replicate
spread reported throughout. As a robustness check on the decoding
strategy we ran the full Direct-mode self-play grid, 4 models by 4
games at 50 rounds, with greedy decoding ($\tau = 0$). Greedy generation
is deterministic, repeated runs produce identical action sequences, so a
single run per cell is exact rather than an estimate, and there were
zero parse failures in all trajectories. The BoS and MP cells reported
here were re-collected under the role-perspective rendering of
Section~\ref{sec:protocol}; every value except one is identical to the
previously reported one, and the exception moves by $0.018$.
Table~\ref{tab:temp0} reports the final Nash distance per cell.

\begin{table}[h]
\centering
\caption{Greedy-decoding ($\tau = 0$) self-play, Direct mode, one
deterministic run per cell, final Nash distance $d^{(50)}$. The headline
split of Table~\ref{tab:behavioral} reproduces cell for cell.}
\label{tab:temp0}
\begin{tabular}{lcccc}
\toprule
Game & L-8B & L-70B & Q-32B & Q-72B \\
\midrule
PD  & 0.040 & 2.000 & 2.000 & 2.000 \\
BoS & 0.000 & 0.000 & 0.333 & 0.000 \\
SH  & 0.000 & 0.000 & 0.000 & 0.000 \\
MP  & 0.762 & 0.000 & 0.000 & 0.000 \\
\bottomrule
\end{tabular}
\end{table}

Three cells deserve their round-level stories, because they are the
clearest illustrations in our data of what the distance metric hides.
First, the three larger models in MP settle into perfectly synchronized
alternation in which Player A wins every single round, cumulative payoff
$+50/-50$ for Llama-70B and Qwen-72B and $+42/-42$ for Qwen-32B. The
per-agent marginals are exactly 50/50, so $d = 0.000$, yet the joint
play is maximally exploitative, and under the corrected rendering Player
B is explicitly informed that it should mismatch and still loses every
round. Greedy decoding cannot implement a mixed strategy at all, so the
MP equilibrium is unreachable by construction at $\tau = 0$, and the
synchronized alternation is the deterministic surrogate the models
adopt, which is the structural reason the main experiments use sampled
decoding. Second, Llama-3-8B in MP begins in the same synchronized
alternation and desynchronizes into a fixed exploited pattern; in the
re-collected run it is Player B that freezes, on Tails, while A keeps
losing into it, cumulative $-24/+24$ and $d = 0.762$, a role-reversed
version of the previously reported desynchronization. Third, Qwen-32B in
BoS alternates between the two equilibria perfectly out of phase with
its identical partner, mismatching on all 50 rounds for a total payoff
of exactly $0.0$ each at $d = 0.333$. This cell reproduces exactly under
the corrected rendering, so it is genuine model behavior rather than a
rendering artifact; at $\tau = 0.7$ the same cell coordinates.
Everywhere else the greedy picture matches the main grid, with all four
models on Stag in SH and three of four locked on Opera in BoS, and the
central PD split is unchanged under deterministic decoding.

\section{Model Outputs Under Intervention}
\label{app:intervention}

The steering and clamping quantities of Section~\ref{sec:causal} are
probabilities renormalized over the two action tokens at the decision
position, properties of the decision readout rather than of free-running
generation. This appendix documents generation directly, with greedy
8-token continuations from the committed PD prompts under the same saved
vector, steering applied at layers 20, 22, and 24 at all positions and
clamping at the final position of layer 22, exactly as in the main
experiments. The readouts measured in the same runs agree with the
Section~\ref{sec:causal} sweep values within $0.02$ on this prompt
subset, so the samples and the quantitative figures share one stimulus
provenance.

\begin{table}[h]
\centering
\small
\caption{Greedy 8-token continuations under intervention on the
committed prompts (two prompts per condition produced identical or
near-identical text; one shown). Text is verbatim except that newlines
are shown as spaces and ``[non-ASCII $\times 8$]'' abbreviates a
repeated non-Latin token. The readout column is the renormalized
$P(\text{Cooperate})$ measured in the same run.}
\label{tab:intervention_samples}
\begin{tabular}{llcl}
\toprule
Intervention & Value & $P(\text{Coop})$ & Greedy continuation \\
\midrule
Steering & $\alpha=-20$ & 0.042 & [non-ASCII $\times 8$] \\
Steering & $\alpha=-5$  & 0.005 & \texttt{.\ My response: (to be filled} \\
None     & $\alpha=0$   & 0.376 & \texttt{Please respond with your answer. I} \\
Steering & $\alpha=+10$ & 0.818 & \texttt{BUT BE HUB CO} (degraded) \\
Steering & $\alpha=+40$ & 0.953 & [non-ASCII $\times 8$] \\
\midrule
Clamping & $c=-30$ & 0.000 & \texttt{.\ My (r)response: (} \\
None     & $c=0$   & 0.389 & \texttt{Please respond with your answer. I} \\
Clamping & $c=+30$ & 0.926 & \texttt{Please make a decision, and we} \\
\bottomrule
\end{tabular}
\end{table}

The pattern is systematic and tracks the structure of the two
interventions. Clamping, which modifies only the final position,
preserves fluent and well-formed text across the entire swept range,
including at the cooperative extreme, where the readout flips decisively
to cooperation while the continuation remains ordinary instructional
English. Steering, which injects at every sequence position, stays
fluent at the moderate defect-side operating point, degrades free-form
fluency at $\alpha = +10$, and collapses it at the swept extremes, even
as the decision-position readout remains sharp and directionally
consistent throughout. This is the expected signature of an all-position
injection, since a large added vector perturbs the representation of the
entire context while the relative ordering of the two action tokens at
the decision position, precisely the axis the direction encodes, remains
reliable. Two scope conclusions follow. The quantitative steering and
clamping claims are claims about the decision readout as defined in
Section~\ref{sec:measurement}, and causal control with fully fluent
generation is demonstrated by the clamping intervention, so the extreme
ends of the steering sweep should be read as stress tests of the
direction rather than as operating points for deployed generation.

\section{Roster and Per-Game Results for the Random Neutral Games}
\label{app:randomgames}

The 48-game roster of Section~\ref{sec:randomgames} is generated
deterministically from roster master seed 20260820 by a committed
generator,
with integer payoffs, per-game randomized assignment of the Circle and
Square labels to the action indices, and stratified sampling over the
five generic structural classes. Every equilibrium is computed by
exhaustive enumeration and independently cross-checked with a standard
solver before any game enters the roster. Each of the 192 model-game
cells runs once under greedy decoding and three times under sampling at
$\tau = 0.7$, 50 rounds per run, on the neutral surface template of
Appendix~\ref{app:prompts}, and the full run records are available from
the authors upon reasonable request.

Tables~\ref{tab:roster_SD2}--\ref{tab:roster_X} list every game with its
bimatrix, its exact equilibria, and each model's Nash distance under
both decoding regimes. Reading them alongside
Section~\ref{sec:randomgames}, the sampled SDs show that most per-game
results are individually stable, with a median within-game SD of
$0.065$, so the headline percentages rest largely on reproducible cells,
while roughly one cell in six has an SD above $0.2$, and the knife-edge
games discussed in the main text sit in that variable minority.

\begin{table}[p]
\centering
\scriptsize
\caption{Stratum SD2 (social dilemmas, both players dominant, equilibrium Pareto-dominated). Payoffs list the bimatrix cells
$(a,b)$ in the order (0,0)\,(0,1)\,/\,(1,0)\,(1,1) over Player A's row
action and Player B's column action; equilibria list pure profiles as
action-index pairs and the mixed profile, where present, as each
player's probability of action 0. Model columns give the sampled
$\tau=0.7$ Nash distance, mean $\pm$ SD over three replicates, with the
deterministic greedy value in parentheses.}
\label{tab:roster_SD2}
\setlength{\tabcolsep}{3.0pt}
\begin{tabular}{llccccc}
\toprule
Game & Payoffs & Equilibria & L-8B & L-70B & Q-32B & Q-72B \\
\midrule
rg01 & (5,6)\,(7,2)\,/\,(4,8)\,(6,7) & (0,0) & 1.92$\pm$0.03 (1.96) & 1.12$\pm$0.06 (1.11) & 0.43$\pm$0.13 (0.20) & 0.08$\pm$0.01 (0.03) \\
rg02 & (0,8)\,(2,5)\,/\,(1,3)\,(8,0) & (1,0) & 0.19$\pm$0.14 (0.04) & 0.79$\pm$0.06 (0.89) & 0.57$\pm$0.05 (0.53) & 0.15$\pm$0.03 (0.06) \\
rg03 & (9,0)\,(5,1)\,/\,(7,7)\,(3,8) & (0,1) & 0.19$\pm$0.12 (0.04) & 0.51$\pm$0.20 (0.81) & 0.62$\pm$0.10 (0.52) & 0.41$\pm$0.23 (1.41) \\
rg04 & (1,3)\,(9,1)\,/\,(0,9)\,(4,8) & (0,0) & 1.40$\pm$0.14 (1.96) & 1.05$\pm$0.02 (1.18) & 0.69$\pm$0.12 (0.64) & 0.20$\pm$0.02 (0.20) \\
rg05 & (8,0)\,(6,3)\,/\,(7,6)\,(4,9) & (0,1) & 0.98$\pm$0.58 (0.06) & 0.47$\pm$0.11 (0.71) & 0.43$\pm$0.09 (0.42) & 0.17$\pm$0.05 (0.03) \\
rg06 & (5,6)\,(3,7)\,/\,(4,8)\,(0,9) & (0,1) & 0.36$\pm$0.04 (0.09) & 0.49$\pm$0.13 (0.46) & 0.41$\pm$0.06 (0.21) & 0.45$\pm$0.32 (1.96) \\
\bottomrule
\end{tabular}
\end{table}

\begin{table}[p]
\centering
\scriptsize
\caption{Stratum D2 (other double-dominance games). Payoffs list the bimatrix cells
$(a,b)$ in the order (0,0)\,(0,1)\,/\,(1,0)\,(1,1) over Player A's row
action and Player B's column action; equilibria list pure profiles as
action-index pairs and the mixed profile, where present, as each
player's probability of action 0. Model columns give the sampled
$\tau=0.7$ Nash distance, mean $\pm$ SD over three replicates, with the
deterministic greedy value in parentheses.}
\label{tab:roster_D2}
\setlength{\tabcolsep}{3.0pt}
\begin{tabular}{llccccc}
\toprule
Game & Payoffs & Equilibria & L-8B & L-70B & Q-32B & Q-72B \\
\midrule
rg07 & (8,5)\,(2,9)\,/\,(3,3)\,(0,5) & (0,1) & 0.36$\pm$0.35 (0.03) & 0.34$\pm$0.13 (0.24) & 0.37$\pm$0.08 (0.48) & 0.10$\pm$0.03 (0.06) \\
rg08 & (1,7)\,(0,9)\,/\,(8,7)\,(9,8) & (1,1) & 0.85$\pm$0.62 (0.08) & 0.53$\pm$0.10 (0.63) & 0.06$\pm$0.00 (0.06) & 0.05$\pm$0.04 (0.00) \\
rg09 & (1,0)\,(4,5)\,/\,(8,0)\,(9,6) & (1,1) & 0.07$\pm$0.03 (0.04) & 0.68$\pm$0.08 (0.64) & 0.00$\pm$0.00 (0.00) & 0.00$\pm$0.00 (0.00) \\
rg10 & (6,1)\,(8,9)\,/\,(0,1)\,(2,4) & (0,1) & 0.14$\pm$0.03 (0.03) & 0.22$\pm$0.03 (0.22) & 0.11$\pm$0.02 (0.06) & 0.02$\pm$0.02 (0.00) \\
rg11 & (3,2)\,(2,5)\,/\,(7,0)\,(8,1) & (1,1) & 0.60$\pm$0.41 (1.39) & 0.66$\pm$0.09 (0.92) & 0.29$\pm$0.07 (0.20) & 0.08$\pm$0.03 (0.00) \\
rg12 & (3,1)\,(4,2)\,/\,(6,2)\,(8,4) & (1,1) & 0.10$\pm$0.07 (0.04) & 0.37$\pm$0.05 (0.43) & 0.00$\pm$0.00 (0.00) & 0.01$\pm$0.02 (0.00) \\
\bottomrule
\end{tabular}
\end{table}

\begin{table}[p]
\centering
\scriptsize
\caption{Stratum D1 (single-dominance games). Payoffs list the bimatrix cells
$(a,b)$ in the order (0,0)\,(0,1)\,/\,(1,0)\,(1,1) over Player A's row
action and Player B's column action; equilibria list pure profiles as
action-index pairs and the mixed profile, where present, as each
player's probability of action 0. Model columns give the sampled
$\tau=0.7$ Nash distance, mean $\pm$ SD over three replicates, with the
deterministic greedy value in parentheses.}
\label{tab:roster_D1}
\setlength{\tabcolsep}{3.0pt}
\begin{tabular}{llccccc}
\toprule
Game & Payoffs & Equilibria & L-8B & L-70B & Q-32B & Q-72B \\
\midrule
rg13 & (0,6)\,(0,1)\,/\,(2,0)\,(3,4) & (1,1) & 0.25$\pm$0.14 (0.17) & 0.56$\pm$0.09 (0.74) & 0.02$\pm$0.04 (0.00) & 0.02$\pm$0.02 (0.03) \\
rg14 & (1,3)\,(0,2)\,/\,(5,1)\,(6,5) & (1,1) & 0.33$\pm$0.25 (0.06) & 0.61$\pm$0.08 (0.66) & 0.01$\pm$0.02 (0.00) & 0.04$\pm$0.02 (0.03) \\
rg15 & (7,1)\,(3,0)\,/\,(3,8)\,(9,3) & (0,0) & 1.96$\pm$0.00 (1.98) & 1.02$\pm$0.04 (1.12) & 0.90$\pm$0.18 (0.88) & 0.32$\pm$0.23 (0.13) \\
rg16 & (7,6)\,(7,8)\,/\,(3,5)\,(1,1) & (0,1) & 0.58$\pm$0.84 (0.03) & 0.20$\pm$0.03 (0.14) & 0.23$\pm$0.05 (0.20) & 0.00$\pm$0.00 (0.00) \\
rg17 & (0,8)\,(4,9)\,/\,(8,2)\,(3,3) & (0,1) & 1.27$\pm$0.22 (1.39) & 1.20$\pm$0.16 (1.20) & 1.26$\pm$0.02 (1.29) & 0.79$\pm$0.61 (0.40) \\
rg18 & (1,5)\,(0,2)\,/\,(2,1)\,(5,6) & (1,1) & 0.86$\pm$0.70 (1.39) & 0.68$\pm$0.17 (0.46) & 0.01$\pm$0.02 (0.00) & 0.01$\pm$0.02 (0.03) \\
rg19 & (9,7)\,(2,8)\,/\,(4,8)\,(0,3) & (0,1) & 0.85$\pm$0.58 (0.29) & 0.46$\pm$0.07 (0.70) & 1.25$\pm$0.11 (1.33) & 0.91$\pm$0.20 (1.41) \\
rg20 & (3,2)\,(8,6)\,/\,(2,6)\,(0,3) & (0,1) & 0.22$\pm$0.05 (0.03) & 0.26$\pm$0.05 (0.26) & 0.30$\pm$0.01 (0.18) & 0.03$\pm$0.03 (0.00) \\
rg21 & (6,5)\,(7,9)\,/\,(3,6)\,(4,2) & (0,1) & 0.36$\pm$0.13 (0.10) & 0.28$\pm$0.09 (0.24) & 0.47$\pm$0.04 (0.40) & 0.11$\pm$0.08 (0.00) \\
rg22 & (9,5)\,(7,0)\,/\,(4,3)\,(5,7) & (0,0) & 0.21$\pm$0.12 (0.06) & 0.35$\pm$0.01 (0.18) & 0.25$\pm$0.11 (0.16) & 0.06$\pm$0.02 (0.06) \\
rg23 & (5,5)\,(0,3)\,/\,(8,0)\,(5,8) & (1,1) & 0.26$\pm$0.22 (2.00) & 0.75$\pm$0.22 (0.50) & 0.12$\pm$0.02 (0.09) & 0.25$\pm$0.18 (0.58) \\
rg24 & (5,7)\,(6,9)\,/\,(9,9)\,(7,5) & (1,0) & 1.10$\pm$0.49 (0.14) & 0.48$\pm$0.08 (0.47) & 0.27$\pm$0.03 (0.18) & 0.06$\pm$0.03 (0.00) \\
\bottomrule
\end{tabular}
\end{table}

\begin{table}[p]
\centering
\scriptsize
\caption{Stratum C (two pure equilibria). Payoffs list the bimatrix cells
$(a,b)$ in the order (0,0)\,(0,1)\,/\,(1,0)\,(1,1) over Player A's row
action and Player B's column action; equilibria list pure profiles as
action-index pairs and the mixed profile, where present, as each
player's probability of action 0. Model columns give the sampled
$\tau=0.7$ Nash distance, mean $\pm$ SD over three replicates, with the
deterministic greedy value in parentheses.}
\label{tab:roster_C}
\setlength{\tabcolsep}{3.0pt}
\begin{tabular}{llccccc}
\toprule
Game & Payoffs & Equilibria & L-8B & L-70B & Q-32B & Q-72B \\
\midrule
rg25 & (4,2)\,(8,7)\,/\,(9,9)\,(3,7) & (0,1) (1,0) m(0.29,0.50) & 0.22$\pm$0.08 (0.78) & 0.44$\pm$0.26 (0.61) & 0.27$\pm$0.10 (0.22) & 0.16$\pm$0.05 (0.72) \\
rg26 & (5,0)\,(7,2)\,/\,(9,6)\,(5,0) & (0,1) (1,0) m(0.75,0.33) & 0.24$\pm$0.03 (1.05) & 0.42$\pm$0.00 (0.42) & 0.57$\pm$0.06 (0.51) & 0.12$\pm$0.02 (0.03) \\
rg27 & (5,6)\,(0,5)\,/\,(0,3)\,(1,7) & (0,0) (1,1) m(0.80,0.17) & 0.07$\pm$0.04 (0.04) & 0.32$\pm$0.20 (0.23) & 0.20$\pm$0.10 (0.13) & 0.22$\pm$0.32 (0.00) \\
rg28 & (8,0)\,(8,8)\,/\,(9,7)\,(2,0) & (0,1) (1,0) m(0.47,0.86) & 0.14$\pm$0.11 (0.04) & 0.40$\pm$0.20 (0.32) & 0.36$\pm$0.04 (0.26) & 0.10$\pm$0.04 (0.00) \\
rg29 & (4,5)\,(7,7)\,/\,(8,9)\,(1,8) & (0,1) (1,0) m(0.33,0.60) & 0.37$\pm$0.39 (0.06) & 0.50$\pm$0.19 (0.33) & 0.41$\pm$0.07 (0.34) & 0.15$\pm$0.06 (0.03) \\
rg30 & (5,3)\,(8,4)\,/\,(6,8)\,(6,1) & (0,1) (1,0) m(0.88,0.67) & 0.29$\pm$0.03 (0.06) & 0.59$\pm$0.05 (0.53) & 0.33$\pm$0.10 (0.32) & 0.24$\pm$0.00 (0.12) \\
rg31 & (5,7)\,(1,5)\,/\,(2,3)\,(2,6) & (0,0) (1,1) m(0.60,0.25) & 0.13$\pm$0.01 (0.04) & 0.36$\pm$0.08 (0.09) & 0.25$\pm$0.14 (0.15) & 0.04$\pm$0.01 (0.00) \\
rg32 & (9,7)\,(7,3)\,/\,(8,5)\,(9,6) & (0,0) (1,1) m(0.20,0.67) & 0.09$\pm$0.05 (0.04) & 0.35$\pm$0.05 (0.20) & 0.19$\pm$0.09 (0.16) & 0.06$\pm$0.00 (0.03) \\
rg33 & (1,2)\,(3,4)\,/\,(4,7)\,(2,4) & (0,1) (1,0) m(0.60,0.25) & 0.26$\pm$0.08 (0.04) & 0.50$\pm$0.14 (0.48) & 0.27$\pm$0.00 (0.23) & 0.13$\pm$0.06 (0.92) \\
rg34 & (8,7)\,(1,5)\,/\,(5,1)\,(8,8) & (0,0) (1,1) m(0.78,0.70) & 0.08$\pm$0.03 (0.06) & 0.11$\pm$0.01 (0.08) & 0.06$\pm$0.00 (0.06) & 0.04$\pm$0.02 (0.03) \\
rg35 & (1,2)\,(8,6)\,/\,(6,8)\,(1,5) & (0,1) (1,0) m(0.43,0.58) & 0.60$\pm$0.30 (0.98) & 0.43$\pm$0.22 (0.41) & 0.55$\pm$0.03 (0.58) & 0.10$\pm$0.02 (0.04) \\
rg36 & (9,9)\,(2,0)\,/\,(5,1)\,(4,2) & (0,0) (1,1) m(0.10,0.33) & 0.18$\pm$0.08 (0.08) & 0.50$\pm$0.07 (0.67) & 0.00$\pm$0.00 (0.00) & 0.00$\pm$0.00 (0.00) \\
\bottomrule
\end{tabular}
\end{table}

\begin{table}[p]
\centering
\scriptsize
\caption{Stratum X (unique mixed equilibrium). Payoffs list the bimatrix cells
$(a,b)$ in the order (0,0)\,(0,1)\,/\,(1,0)\,(1,1) over Player A's row
action and Player B's column action; equilibria list pure profiles as
action-index pairs and the mixed profile, where present, as each
player's probability of action 0. Model columns give the sampled
$\tau=0.7$ Nash distance, mean $\pm$ SD over three replicates, with the
deterministic greedy value in parentheses.}
\label{tab:roster_X}
\setlength{\tabcolsep}{3.0pt}
\begin{tabular}{llccccc}
\toprule
Game & Payoffs & Equilibria & L-8B & L-70B & Q-32B & Q-72B \\
\midrule
rg37 & (9,3)\,(1,8)\,/\,(4,4)\,(6,3) & m(0.17,0.50) & 1.06$\pm$0.33 (1.34) & 0.74$\pm$0.03 (0.62) & 0.29$\pm$0.03 (0.31) & 0.11$\pm$0.04 (0.25) \\
rg38 & (6,1)\,(8,8)\,/\,(2,6)\,(9,0) & m(0.46,0.20) & 0.74$\pm$0.10 (0.78) & 0.48$\pm$0.14 (0.43) & 0.65$\pm$0.27 (0.32) & 0.33$\pm$0.25 (0.81) \\
rg39 & (0,9)\,(9,8)\,/\,(9,6)\,(0,8) & m(0.67,0.50) & 0.94$\pm$0.06 (1.12) & 0.56$\pm$0.15 (0.44) & 0.66$\pm$0.04 (0.70) & 0.80$\pm$0.03 (0.83) \\
rg40 & (5,5)\,(3,6)\,/\,(4,7)\,(6,6) & m(0.50,0.75) & 1.04$\pm$0.31 (1.24) & 0.51$\pm$0.04 (0.51) & 0.77$\pm$0.18 (1.06) & 0.54$\pm$0.25 (0.57) \\
rg41 & (2,6)\,(2,8)\,/\,(0,7)\,(8,6) & m(0.33,0.75) & 1.15$\pm$0.02 (1.15) & 0.77$\pm$0.11 (0.94) & 0.54$\pm$0.07 (0.10) & 0.47$\pm$0.20 (1.15) \\
rg42 & (2,0)\,(3,3)\,/\,(0,5)\,(9,4) & m(0.25,0.75) & 1.11$\pm$0.00 (1.11) & 0.63$\pm$0.08 (0.30) & 0.88$\pm$0.06 (1.01) & 1.11$\pm$0.01 (1.07) \\
rg43 & (1,6)\,(4,2)\,/\,(3,3)\,(3,8) & m(0.56,0.33) & 0.86$\pm$0.02 (0.74) & 0.24$\pm$0.06 (0.24) & 0.34$\pm$0.32 (0.04) & 0.42$\pm$0.23 (0.92) \\
rg44 & (0,8)\,(4,5)\,/\,(7,2)\,(3,5) & m(0.50,0.12) & 1.10$\pm$0.11 (1.39) & 0.91$\pm$0.04 (0.94) & 0.19$\pm$0.02 (0.18) & 0.32$\pm$0.09 (0.72) \\
rg45 & (7,1)\,(6,7)\,/\,(6,2)\,(9,1) & m(0.14,0.75) & 1.05$\pm$0.02 (1.07) & 0.53$\pm$0.03 (0.55) & 0.23$\pm$0.08 (0.07) & 0.53$\pm$0.07 (1.03) \\
rg46 & (7,1)\,(0,9)\,/\,(3,5)\,(1,0) & m(0.38,0.20) & 0.86$\pm$0.30 (0.39) & 0.16$\pm$0.05 (0.19) & 0.40$\pm$0.09 (0.38) & 0.28$\pm$0.12 (1.02) \\
rg47 & (1,6)\,(8,3)\,/\,(4,6)\,(0,9) & m(0.50,0.73) & 0.52$\pm$0.12 (1.19) & 0.20$\pm$0.03 (0.13) & 1.04$\pm$0.02 (0.77) & 0.12$\pm$0.09 (0.26) \\
rg48 & (2,8)\,(7,0)\,/\,(5,2)\,(5,7) & m(0.38,0.40) & 0.83$\pm$0.25 (1.22) & 0.57$\pm$0.05 (0.70) & 0.44$\pm$0.05 (0.44) & 0.57$\pm$0.29 (0.76) \\
\bottomrule
\end{tabular}
\end{table}

\section{Steering Robustness and Random-Direction Controls}
\label{app:robustness}

This appendix tabulates the experiments behind
Section~\ref{sec:robustness}. Three fresh mean-difference vectors were
extracted at layer 22 from the three committed extraction sets, sign-fixed
to the cooperative-positive convention, together with their normalized
average and three seeded random unit directions. All vectors are archived
with the results.

\begin{table}[h]
\centering
\small
\caption{Pairwise cosine similarities. The three re-extracted vectors
agree with each other almost perfectly while all are essentially
orthogonal to the archived vector.}
\label{tab:robust_cos}
\begin{tabular}{lccccc}
\toprule
 & saved & extract\_a & extract\_b & extract\_c & average \\
\midrule
saved      & $+1.000$ & $-0.016$ & $-0.011$ & $-0.014$ & $-0.014$ \\
extract\_a &          & $+1.000$ & $+0.979$ & $+0.995$ & $+0.996$ \\
extract\_b &          &          & $+1.000$ & $+0.987$ & $+0.993$ \\
extract\_c &          &          &          & $+1.000$ & $+0.998$ \\
\bottomrule
\end{tabular}
\end{table}

\begin{table}[h]
\centering
\small
\caption{Anchor-point readouts per direction on the committed PD prompts.
The saved vector alone combines strong bidirectional steering with full
clamping effectiveness; the re-extracted directions steer at strength but
are nearly clamp-inert; random directions are erratic and
sign-inconsistent.}
\label{tab:robust_anchors}
\begin{tabular}{lccccc}
\toprule
Direction & $\alpha{=}-5$: $P(\text{D})$ & $\alpha{=}+10$: $P(\text{C})$ &
$\alpha{=}+40$: $P(\text{C})$ & $c{=}-30$: $P(\text{C})$ & $c{=}+30$: $P(\text{C})$ \\
\midrule
saved      & 0.995 & 0.821 & 0.953 & 0.000 & 0.925 \\
extract\_a & 0.959 & 0.120 & 0.996 & 0.048 & 0.081 \\
extract\_b & 0.982 & 0.464 & 0.998 & 0.020 & 0.506 \\
extract\_c & 0.975 & 0.179 & 0.997 & 0.034 & 0.206 \\
average    & 0.973 & 0.229 & 0.997 & 0.037 & 0.262 \\
random\_0  & 0.261 & 0.006 & 0.001 & 0.786 & 0.111 \\
random\_1  & 0.872 & 0.684 & 0.157 & 0.039 & 0.064 \\
random\_2  & 0.434 & 0.188 & 0.100 & 0.858 & 0.038 \\
\bottomrule
\end{tabular}
\end{table}

\section{Protocol Unification Notes}
\label{app:protocol}

Every mechanistic and interventional number in this version is produced
under one protocol, defined in Section~\ref{sec:measurement}, on the
committed seeded stimulus sets, and gated before interpretation. The
anchor gate loads the archived vector and the committed PD set and
requires the recorded behavior to reproduce, which it does, with the
lens at $0.952$ at layer 30 against the prior fresh-sample $0.951$,
final $0.393$ in both, and all steering and clamping anchors within a
point. Every lens run additionally asserts that its final-layer value
equals the model's true output readout on the same prompts.

Earlier versions of four pipelines differed from this protocol, and we
record the differences so that no number has to be taken on trust. The
earlier version's 8B lens figure drew its stimuli from an unseeded
generator, and
its peak of about $0.80$ reflects that draw; two independent seeded
samples agree within $0.012$ at every reported point, and the seeded
values replace it. The published 70B and Qwen-72B lenses used a
script-local prompt builder rather than the shared one and applied the
unembedding without the final normalization on intermediate layers, and
in addition the last entry of the HuggingFace hidden-state tuple is
already post-normalization, so a uniform formula double-normalizes the
final point; the unified pipeline handles the indexing correctly,
verifies the final point against the true readout per run, and its
values, with the 70B final at $0.821$ against the earlier $0.452$ under
the old prompts, replace the old ones. The published cross-game steering
script extracted its own vector on its own direct-mode prompts; the
unified experiment applies the archived vector to the committed sets,
which is what the main text claims, and the transfer conclusions are
stated from that experiment. Finally, the earlier version's head-ablation code
hooked the attention-result activation without enabling its computation,
so its hooks never fired and its published values were seven copies of
the baseline; the re-implemented experiment enables the pathway,
verifies it with a whole-layer positive control that must move the
readout, and only then reports the per-head effects. The corrected
behavioral rendering for the two asymmetric games, and the null impact
of that correction, are described in Sections~\ref{sec:protocol}
and~\ref{sec:behavioral}.

\end{document}